\documentclass[aip,cha,reprint]{revtex4-1}

\usepackage{amsmath,amsfonts}
\usepackage{bm}
\usepackage{graphicx}
\usepackage{enumerate}

\newtheorem{theorem}{Theorem}

\newlength{\Mylen}

\newcommand{\PGF}{[\nabla F]_{xy}}

\begin{document}

\title{Finite-time barriers to front propagation in two-dimensional
  fluid flows}

\author{John R.~Mahoney}
\email{jmahoney3@ucmerced.edu}

\author{Kevin A.~Mitchell}
\email{kmitchell@ucmerced.edu}

\affiliation{School of Natural Sciences, University of California,
Merced, California, 95343}

\date{\today}

\begin{abstract}
  Recent theoretical and experimental investigations have demonstrated
  the role of certain invariant manifolds, termed \emph{burning
    invariant manifolds} (BIMs), as one-way dynamical barriers to
  reaction fronts propagating within a flowing fluid.  These barriers
  form one-dimensional curves in a two-dimensional fluid flow.  In
  prior studies, the fluid velocity field was required to be either
  time-independent or time-periodic.  In the present study, we develop
  an approach to identify prominent one-way barriers based only on
  fluid velocity data over a finite time interval, which may have
  arbitrary time-dependence.  We call such a barrier a \emph{burning
    Lagrangian coherent structure} (bLCS) in analogy to Lagrangian
  coherent structures (LCSs) commonly used in passive advection.  Our
  approach is based on the variational formulation of LCSs using
  curves of stationary ``Lagrangian shear'', introduced by Farazmand,
  Blazevski, and Haller [Physica D 278-279, 44 (2014)] in the context
  of passive advection.  We numerically validate our technique by
  demonstrating that the bLCS closely tracks the BIM for a
  time-independent, double-vortex channel flow with an opposing
  ``wind''.
\end{abstract} 

\maketitle

\begin{quotation}
  Looking into a pocket near the edge of a lazy stream, one may see
  froth and small plant particulates twisting and evolving in sinuous
  strands.  While streams are always subject to change---a branch
  falls in, a rock rolls, a fish swims nearby---the observed pattern
  changes little due to these differences in the stream's current.  An
  explanation for this is that the pattern organizes around a robust
  set of curves that form a ``skeleton'' for particulate advection.

  While it is clear to the eye that such curves exist, it is another
  thing to define them precisely. A variety of techniques have been
  developed for this purpose, and while the original techniques were
  limited to flows that were either constant or periodic in time, the
  most general modern techniques can analyze flows with arbitrary
  time-dependence.

  Considering now not a stream, but the entire ocean, one can find a
  teaming algae population within large eddies.  The ``living
  pattern'' created by the swirling collection of replicating
  organisms is different from the passive particulates because the
  algae cluster is active---replication increases the size of the
  algae patch.  There is a corresponding new set of ``active'' curves
  underlying the pattern formed by this growing front.

  In this paper, we develop a technique to extract the most attracting
  and repelling curves for such active material within an arbitrary
  flow defined over a finite time interval. We imagine eventual
  application to improving management of oceanic algae blooms or to
  informing the design of combustion systems.
\end{quotation}

\section{Introduction}
\label{sec:intro}

A growing interest has recently developed in fluid systems that
combine the phenomenon of (chaotic) advection with that of front
propagation.  This interest is due in part to geophysical and
technological applications, such as the growth of plankton blooms in
oceanic flows~\cite{Neufeld09,Scotti07,Sandulescu08}, the growth of
the ozone hole in the atmosphere, and the spreading of chemical
reactions in microfluidic devices.  Another important driver of this
field has been a steady accumulation of experimental observations and
measurements of (Belousov-Zhabotinsky) chemical reaction fronts
spreading in driven, laboratory-scale flows
\cite{Paoletti05,Paoletti05b,Paoletti06,Schwartz08,Boehmer08,OMalley11,Pocheau06,Pocheau08,vonKameke13},
as well as early theoretical predictions on such
advection-reaction-diffusion systems \cite{Abel01,Abel02,Cencini03}.

A recent advance in understanding advection-reaction-diffusion (ARD)
systems is the realization that invariant manifold theory, familiar in
the analysis of passive advective
mixing~\cite{MacKay84,Ottino89,Rom-Kedar90b,Wiggins92}, is also
applicable to the analysis of front propagation in
fluids~\cite{Mahoney12,Mitchell12b,Bargteil12,Mahoney13,Mahoney15b,Gowen14}.
Such manifolds have been called \emph{burning invariant manifolds}, or
BIMs, to distinguish them from their analogous, but distinct,
advective counterparts.  The most important property of BIMs is that
they form time-invariant, or time-periodic, one-way barriers to front
propagation in fluid flows that are themselves either time-invariant
or time-periodic, respectively.  The unstable BIMs are attracting
structures, in that an initial stimulation near such a BIM ultimately
creates a reaction front in the fluid that converges upon the BIM.  In
the most striking cases, this front persists for arbitrarily long
times, forming a frozen or pinned front, whose profile follows the
BIM~\cite{Mahoney15b}.  Alternatively, stable BIMs are repelling
structures.  Initial stimulation points on either side have
qualitatively distinct future behavior.  Various other aspects of
passive invariant manifolds extend to BIMs.  For example, BIMs lead to
a theory of turnstiles for front propagation in time-periodic fluid
flows~\cite{Mahoney13}.

Despite the successful application of BIMs to ARD systems, their
applicability has been limited to flows that are either
time-independent or time-periodic.  The present paper addresses this
limitation by developing a theory of prominent, one-way barriers to
front propagation in general fluid flows specified over a finite
interval of time, with no assumption on the time-dependence over this
interval.  These barriers, called \emph{burning Lagrangian coherent
  structures} (bLCSs), are either locally most repelling or most
attracting.  We validate this theory by numerically demonstrating that
the bLCS reproduces the BIMs for time-independent flows considered
over a sufficiently long time interval.  A future publication will
assess this theory in the general case of unsteady flows.

Our work is motivated by recent progress in the use of finite-time
Lyapunov exponent (FTLE) fields~\cite{Pierrehumbert93,
  vonHardenberg00, Doerner99, Haller01, Haller02} and the associated
Lagrangian coherent structures (LCSs) for passive advection in
unsteady
flows~\cite{Haller00a,Haller00b,Haller01,Haller02,Haller11,Shadden05,Shadden06}.
A common approach in advection studies is to assume that ridges of the
(forward) FTLE field are (repelling) LCSs that approximately separate
trajectories with different future behavior.  The FTLE ridge approach
has its deficiencies, however, and can produce both false negatives
and false positives when identifying LCSs~\cite{Haller11}, so it is
best used, perhaps, as an intuitive diagnostic tool.  That said, if
one were to implement the FTLE-ridge approach for front-propagation
dynamics, one would immediately encounter a challenge of dimensions.
This is because the front-element dynamics in two spatial dimensions
is naturally represented as a three-dimensional dynamical system, with
$x$ and $y$ the two spatial coordinates and $\theta$ the orientation
of the front element.  (See Sect.~\ref{sec:preliminaries}.)  Note that
the objects of our interest, the bLCSs, are still one-dimensional
curves, just like the BIMs they seek to generalize; that is, a bLCS is
not a codimension-one object in our study.  Thus, a possible
implementation of the FTLE-ridge approach to the front-element
dynamics would not seek to follow 1D ridges in a 2D space, or even 2D
ridges in a 3D space, but rather to follow 1D ridges in a 3D space.
We see no obvious reason why such curves in the 3D phase space would
correspond to fronts in the $xy$-position space, since a curve in the
3D space must satisfy the front-compatibility criterion (see
Sect.~\ref{sec:preliminaries}) for it to be the lift of a front in
$xy$-space.  It remains an open question, then, whether the 1D FTLE
ridges would even yield fronts in the $xy$-position space, and
assuming they did, these would undoubtedly suffer the same
deficiencies noted in Ref.~\cite{Haller11} for passive advection.

Instead of seeking FTLE ridges, we base our approach on the recent
work of Farazmand, Blazevski, and Haller (FBH)~\cite{Farazmand14},
which uses a variational principle to define the LCS.  The advantage
of a variational formulation is that it is based from the start on a
search for curves in $xy$-space.  These curves could be material
lines, as in FBH~\cite{Farazmand14}, or they could be fronts, as in
the current study.  The fact that the evolution of each front element
depends on its orientation $\theta$ is naturally accommodated within
the variational approach.  Following FBH~\cite{Farazmand14}, we base
the variational integral on the Lagrangian shear, generalized to the
front propagation dynamics.  The output of the variational approach,
when combined with the front compatibility criterion, is a flow on a
2D surface, which we call the \emph{shearless surface}, embedded
within the 3D phase space.  This flow foliates the shearless surface
into \emph{shearless fronts}, which are candidates for a bLCS.  This
reduction to a 2D surface tames the challenge of dimensions discussed
above.  One can, for example, plot burning FTLE (bFTLE) fields on the
2D surface.  However, the dimensional challenges have not been
completely eliminated, since the shearless surface can have a
complicated geometry within the 3D phase space, with folds and
intersections that are not easily represented in a 2D projection.
This geometry raises new considerations in the search for bLCSs that
do not arise when searching for LCSs in passive advection.

This article has the following structure.
Section~\ref{sec:preliminaries} reviews our prior dynamical systems
approach to front propagation in fluid flows and introduces the
concept of BIMs as one-way barriers.  Section~\ref{sec:CoreAnalysis}
contains the core derivations.  The main result is
Theorem~\ref{thm:ShearlessFronts} of Sect.~\ref{sec:ShearlessFronts},
which identifies perfect shearless fronts as the candidates for a
bLCS.  Sect.~\ref{sec:computingSF} describes how to compute these
shearless fronts, and Sect.~\ref{sec:NormalRepulsion} formulates the
selection of the (repelling) bLCS from the set of shearless fronts as
finding that front which maximizes the average normal repulsion.
Section~\ref{sec:DoubleVortex} applies the bLCS approach to a
time-independent double-vortex in a channel with wind.  We study both
short (Sect.~\ref{sec:short}) and long (Sect.~\ref{sec:long})
evolution times to show that the bLCS closely reproduces the BIM of
the time-independent flow. 

\section{Preliminaries}
\label{sec:preliminaries}

\begin{figure}
\centering
\includegraphics[width=\linewidth]{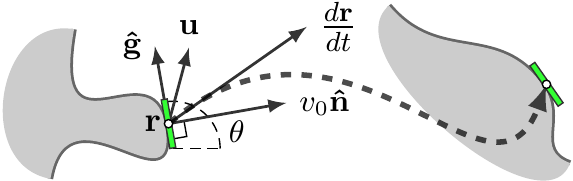}
\caption{
 (color online) Each front element evolves according to
  Eqs.~(\ref{eq:3DODE}).}
\label{fig:3DODE}
\end{figure}

Following previous
studies~\cite{Mahoney12,Mitchell12b,Bargteil12,Mahoney13,Mahoney15b,Gowen14},
we give a brief account of front propagation from the dynamical
systems perspective.  We make the sharp front assumption, in which the
front marks a clear delineation between a reacted or ``burned'' region
and an ``unburned'' region.  Then, a front in a two-dimensional fluid
flow can be represented as a curve $R(s) = (\mathbf{r}(s),\theta(s))$ in the
three-dimensional $xy\theta$-space, where $s$ is an arbitrary smooth
parameterization.  Here $\theta$ denotes the orientation of the normal
$\hat{\mathbf{n}} = (\sin \theta, -\cos \theta)$ and tangent
$\hat{\mathbf{g}} = (\cos \theta, \sin \theta)$ to the front, where
$\hat{\mathbf{n}} \times \hat{\mathbf{g}} = +1$.
(Fig.~\ref{fig:3DODE}.)  The normal vector points in the propagation,
or burning, direction of the front.

Not all curves in $xy\theta$-space represent fronts.  There must be
pointwise agreement between the tangent vector $\mathbf{r}' = d
\mathbf{r}/d s$ of the curve $\mathbf{r}(s)$ and the orientation
$\hat{\mathbf{g}}(\theta)$ given by $\theta$.  We call this the
\emph{front-compatibility criterion}
\begin{equation}
\frac{ \mathbf{r}'}{| \mathbf{r}'|} = \pm \hat{\mathbf{g}} 
= \pm(\cos \theta, \sin \theta).
\label{r26}
\end{equation}

The front propagation speed $v_0$ in the local fluid frame, which we
call the \emph{burning speed}, is assumed to be isotropic and
homogeneous throughout the fluid.  Furthermore, we assume that $v_0$
is independent of the local curvature of the front or any other front
property, i.e. $v_0$ is entirely constant in our analysis.  Then each
front element $(\mathbf{r},\theta)$ evolves independently of its
neighboring front elements according to (see Fig.~\ref{fig:3DODE})
\begin{subequations}
\begin{align}
\frac{d\mathbf{r}}{dt} &= \mathbf{u} + v_0 \hat{\mathbf{n}}, \label{eq:3DODEa} \\ 
\frac{d\theta}{dt} &= - \hat{n}_i u_{i,j} \hat{g}_j, \label{eq:3DODEb}
\end{align}
\label{eq:3DODE}
\end{subequations}
where $\mathbf{u}=\mathbf{u}(\mathbf{r},t)$ is the fluid velocity
field, dot denotes the time derivative, comma denotes differentiation
with respect to $r_i$, and repeated indices are summed.  The fixed
points of this dynamical system are called burning fixed points.  For
time-independent or time-periodic fluid velocity fields
$\mathbf{u}(\mathbf{r},t)$, burning fixed points may have any
combination of stable (S) and unstable (U) directions: SSS, SSU, SUU,
UUU.  The one-dimensional stable and unstable manifolds of SUU and SSU
burning fixed points, respectively, play a special role and are called
burning invariant manifolds (BIMs).  BIMs satisfy the
front-compatibility criterion and may thus be viewed as (virtual)
fronts within the fluid.  As such, they form one-way barriers to front
propagation, owing to the fact that no front can catch up to and
surpass another front moving in the same direction (what we call the
no-passing lemma~\cite{Mitchell12b}.)  BIMs are thus natural,
dynamically defined, geometric objects that restrict the propagation
of fronts in an advecting fluid flow that is either time-independent
or time-periodic.

\section{Coherent structures as solutions to an optimization problem}

\label{sec:CoreAnalysis}

\subsection{Lagrangian shear for fronts}

\begin{figure}
\centering
\includegraphics[width=\linewidth]{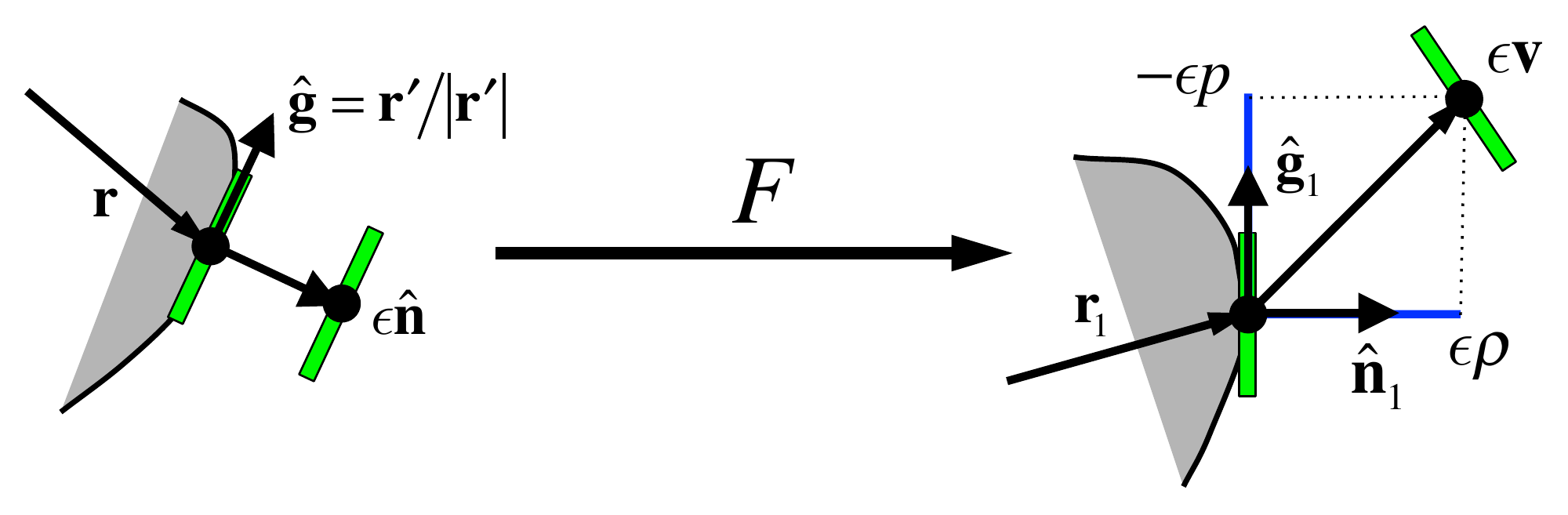}
\caption{\label{fig:LagrangianShear} (color online) The definition of
  Lagrangian shear $p$ and normal repulsion $\rho$.  An initial normal
  displacement $\epsilon \mathbf{\hat{n}}$ of a front element (left)
  evolves under $F$ into a displacement $\epsilon \mathbf{v}$ whose
  tangential projection (blue) is $-\epsilon$ times the Lagrangian
  shear $p$.  The normal projection (blue) is $\epsilon$ times the
  normal repulsion.}
\end{figure}

We adapt the definition of Lagrangian shear in FBH~\cite{Farazmand14}
to the front propagation scenario.  We first assume that
$\mathbf{u}(\mathbf{r},t)$ is given between an initial time $t_0$ and
a final time $t_1 = t_0 + T$, with no requirement that the flow be
time-independent or periodic.  Then the evolution of a front element
over this time interval is obtained by solving Eqs.~(\ref{eq:3DODE}).
We denote the flow map acting on the 3D phase space by $F =
F_{t_0}^{t_1}$.  Consider now an initial front $R(s) =
(\mathbf{r}(s),\theta(s))$ at time $t_0$, and suppose the element at a
given position $\mathbf{r}(s)$ is displaced in the normal direction an
infinitesimal distance $\epsilon \mathbf{\hat{n}}$.  (The orientation
$\theta$ of the front element is not perturbed.)  Under the time
evolution $F$, the point $R = (\mathbf{r},\theta)$ maps forward to
$F(R) = R_1 = (\mathbf{r}_1,\theta_1)$, and the relative displacement
$\epsilon \hat{\mathbf{n}}$ between the original and perturbed front
elements maps forward to a vector $\epsilon \mathbf{v}$ that will in
general have a nonzero tangential projection onto the time-evolved
front.  (See Fig.~\ref{fig:LagrangianShear}.)  We define the
Lagrangian shear $p$ over the time interval $[t_0, t_1]$ to be this
projected length divided by $\epsilon$.

To compute $p$, we first note that the 3D normal displacement vector
$[\hat{\mathbf{n}}, 0]$ at $t_0$ evolves forward to
$\nabla F [\hat{\mathbf{n}}, 0]$ at $t_1$, where $\nabla F$ is the
$3 \times 3$ gradient of $F$, with components
$(\nabla F)_{ij} = F_{i,j}$.  Since $\mathbf{v}$, shown in
Fig.~\ref{fig:LagrangianShear}, is the $xy$-component of
$\nabla F [\hat{\mathbf{n}}, 0]$, we have
\begin{equation}
\mathbf{v} = \Pi_{xy}  \nabla F \Pi_{xy}^T \hat{\mathbf{n}},
\label{r30}
\end{equation}
where
\begin{equation}
\Pi_{xy}
= \left(
\begin{array}{ccc}
1 & 0 & 0 \\
0 & 1 & 0 
\end{array}
\right) 
\end{equation}
is the restricted projection operator from the $xy\theta$-tangent
space into the $xy$-tangent plane.  Note $[\hat{\mathbf{n}}, 0] =
\Pi_{xy}^T \hat{\mathbf{n}}$.  We then define
\begin{equation}
\PGF  = \Pi_{xy}  \nabla F \Pi_{xy}^T 
\end{equation}
to be the $2 \times 2$ restriction of $\nabla F$ to the $xy$-tangent
plane.  Note also that Eq.~(\ref{r26}) implies
\begin{equation}
\hat{\mathbf{n}}
= -\Omega \hat{\mathbf{g}}
=\mp \Omega \mathbf{r}'/|\mathbf{r}'|,
\end{equation}
where 
\begin{equation}
\Omega = 
\left(
\begin{array}{cc}
0 & -1  \\
1 & 0  \\
\end{array}
\right). 
\end{equation}
Then Eq.~(\ref{r30}) becomes
\begin{align}
\mathbf{v} &= \PGF \hat{\mathbf{n}} \label{r31} \\
&= \mp \PGF \Omega \mathbf{r}'/|\mathbf{r}'|.
\label{r10}
\end{align}

We next need an expression for the tangent vector $\hat{\mathbf{g}}_1$
to the front at $\mathbf{r}_1$.  The 3D tangent vector to the curve
$R(s)$ at time $t_0$ is of the form $[\hat{\mathbf{g}}, \alpha]$.
This vector maps forward under $\nabla F$ to a vector whose
$xy$-component $\Pi_{xy} \nabla F (\hat{\mathbf{g}}, \alpha)$ is
proportional to $\hat{\mathbf{g}}_1$.  Note that this fact is actually
true regardless of the value of $\alpha$, since $\alpha$ depends on
the curvature $d \theta / d s$ of $\mathbf{r}(s)$ and not its tangent.
We may thus set $\alpha = 0$ to obtain
\begin{equation}
\hat{\mathbf{g}}_1
 =\frac{\PGF \hat{\mathbf{g}}}{|\PGF  \hat{\mathbf{g}}|}
 =\pm \frac{\PGF  \mathbf{r}'}{| \PGF  \mathbf{r}'|},
\label{r28}
\end{equation}
where we again used the front-compatibility criterion Eq.~(\ref{r26}).
We are now in a position to compute $p$ via
\begin{align}
  p &= \langle -\hat{\mathbf{g}}_1, \mathbf{v} \rangle 
= \frac{ \langle \PGF \mathbf{r}', \PGF \Omega \mathbf{r}' \rangle}
  {| \mathbf{r}' | | \PGF \mathbf{r}' |} \nonumber \\
  &= \frac{ \langle \mathbf{r}', \PGF^T  \PGF \Omega \mathbf{r}' \rangle}
  {\sqrt{ \langle \mathbf{r}', \mathbf{r}' \rangle
\langle \mathbf{r}' ,  \PGF^T  \PGF \mathbf{r}' \rangle }}. 
\label{r2}
\end{align}
The minus sign on $\hat{\mathbf{g}}_1$ conforms to the convention of
FBH~\cite{Farazmand14}.  To summarize, we may express the Lagrangian
shear as
\begin{align}
p(\mathbf{r}, \theta ) = p(\mathbf{r}, \mathbf{r}') &= \frac{\langle \mathbf{r}',  
C(\mathbf{r}, \theta) \Omega \; \mathbf{r}' \rangle}
{\sqrt{ \langle \mathbf{r}',\mathbf{r}' \rangle
\langle \mathbf{r}', C(\mathbf{r}, \theta) \;  \mathbf{r}' \rangle}} \\
&= 
\frac{\langle \mathbf{r}',  D(\mathbf{r}, \theta) \; \mathbf{r}' \rangle}
{\sqrt{ \langle \mathbf{r}',\mathbf{r}' \rangle
\langle \mathbf{r}', C(\mathbf{r}, \theta) \;  \mathbf{r}' \rangle}},
\label{r7}
\end{align}
where 
\begin{align}
C(\mathbf{r}, \theta) &= \PGF^T  \PGF, \label{r14} \\
D(\mathbf{r}, \theta) &=\frac{1}{2}
(C(\mathbf{r}, \theta) \Omega - \Omega C(\mathbf{r}, \theta) ).
\label{r22}
\end{align}
Note that $C$ and $D$ are both $2 \times 2$ symmetric matrices.  We
call $C$ the \emph{projected} Cauchy-Green tensor.  Note that even
though $C$ is a $2 \times 2$ matrix acting on the $xy$-tangent plane,
it still depends on the orientation coordinate $\theta$.  In the limit
that the burning speed $v_0$ vanishes, the projected Cauchy-Green
tensor reduces to the usual Cauchy-Green tensor defined for passive
advection.  Finally, note that we have written $p$ either as a
function of $(\mathbf{r},\theta)$ or $(\mathbf{r},\mathbf{r}')$, since
$p$ does not depend on the magnitude of $\mathbf{r}'$ and since
$\theta$ is the angle of $\mathbf{r}'$, as given by Eq.~(\ref{r26}).

\subsection{Normal repulsion for fronts}

The bLCS will be defined to locally maximize the average normal
repulsion $\rho$.  Adapting the definition of normal repulsion in
Ref.~\cite{Haller11} to the front propagation scenario, we define
$\rho$ to be the normal component of the time-evolved
displacement $\mathbf{v}$ in Fig.~\ref{fig:LagrangianShear}.  That is,
the normal repulsion is a local measure of the degree to which a
nearby front is pushed away, with $\rho > 1$ indicating repulsion and
$\rho<1$ attraction.  To compute $\rho$, we shall need the normal
$\hat{\mathbf{n}}_1$ at the final time $t_1$, which we claim is given
by
\begin{equation}
\hat{\mathbf{n}}_1 = \frac{\PGF^{-1 T}
  \hat{\mathbf{n}}}{|\PGF^{-1 T} \hat{\mathbf{n}}|}.
\label{r11}
\end{equation}
To prove this, we need only show that $\langle \hat{\mathbf{g}}_1,
\hat{\mathbf{n}}_1 \rangle = 0$.  From Eqs.~(\ref{r28}) and (\ref{r11}),
$\langle \hat{\mathbf{g}}_1, \hat{\mathbf{n}}_1 \rangle$ is proportional to
\begin{align}
\langle \PGF \mathbf{r}',
\PGF^{-1 T} \hat{\mathbf{n}} \rangle  &= 
\langle \PGF^{-1} \PGF  \mathbf{r}', \hat{\mathbf{n}} \rangle
\nonumber \\
& = \langle \mathbf{r}' , \hat{\mathbf{n}}  \rangle = 0.
\end{align}
Using Eq.~(\ref{r31}), the normal repulsion is
\begin{align}
\rho & = \langle \hat{\mathbf{n}}_1, \mathbf{v} \rangle 
= \frac{\langle \PGF^{-1 T} \hat{\mathbf{n}}, \PGF \hat{\mathbf{n}} \rangle }
{|\PGF^{-1 T}\hat{\mathbf{n}}|} \nonumber \\
& = \frac{ 1 }
{\sqrt{\langle \PGF^{-1 T} \hat{\mathbf{n}}, \PGF^{-1 T} \hat{\mathbf{n}}\rangle }} 
= \frac{ 1 }
{\sqrt{\langle \hat{\mathbf{n}}, C^{-1} \hat{\mathbf{n}}\rangle }}. 
\end{align}
In summary, the normal repulsion is a function on the 3D phase space
given by 
\begin{equation}
\rho(\mathbf{r},\theta) = 
\rho(\mathbf{r}, \mathbf{r}')=
\frac{1}{\sqrt{\hat{\mathbf{n}} \cdot C^{-1}(\mathbf{r},\theta) \hat{\mathbf{n}}}}.
\label{r29}
\end{equation}

\subsection{Variational problem}
\label{sec:variation}

The variational problem is now stated as follows.  We seek a curve
$\gamma$, parameterized as $\mathbf{r}(s)$, that makes the average
Lagrangian shear
\begin{equation}
  P[\gamma] 
  = \frac{1}{\sigma} \int_{0}^{\sigma}
  p(\mathbf{r}, \mathbf{r}') d s 
\label{r6}
\end{equation}
stationary under infinitesimal variations, where the average is taken
over the interval $s_0 \le s \le s_0 + \sigma$, choosing the initial
point to have parameter $s_0 = 0$.  The endpoints of $\gamma$ are
assumed fixed under the variations.  Note that
$p(\mathbf{r}(s),\mathbf{r}'(s))$ does not depend on how the curve
$\gamma$ is parameterized, since the numerator and denominator in
Eq.~(\ref{r7}) are both second order in $d/ds$.  However, the average
in Eq.~(\ref{r6}) does depend on the parameterization through the line
element $d s$.  Different choices of the parameterization will change
the weighting of the average along $\gamma$.  The stationarity
condition of $P[\gamma]$ should be understood not only in terms of
variations in $\gamma$ but also in terms of variations of the
parameterization $s$ along $\gamma$, with the interval of integration
$[0, \sigma]$ held constant.  (In contrast, one could use the
euclidean line element $\sqrt{\langle \mathbf{r}', \mathbf{r}'
  \rangle} d s$ for the average, in which case $P[\gamma]$ would be
independent of the parameterization of $\gamma$.)  We shall find it
useful to change the parameterization of the stationary curves, so we
next summarize the reparameterization of trajectories in the
Hamiltonian and Lagrangian formalisms.

\subsection{Trajectory reparameterization for Hamiltonian systems}

Suppose $(\mathbf{R}(s),\mathbf{P}(s))$ is a particular trajectory for
an $s$-independent Hamiltonian $H(\mathbf{r},\mathbf{p})$ and let $E$
be the value of the (conserved) Hamiltonian along the trajectory.
Here $s$ plays the role of time.  Define a new trajectory
parameterization $\tau$ by
\begin{equation}
d \tau/d s = \kappa(\mathbf{r},\mathbf{p}), 
\label{r3}
\end{equation}
for some phase space function $\kappa$, and denote the reparameterized
trajectory by $(\tilde{\mathbf{R}}(\tau),\tilde{\mathbf{P}}(\tau))$.  Then
$(\tilde{\mathbf{R}}(\tau),\tilde{\mathbf{P}}(\tau))$ is a trajectory of the
Hamiltonian
\begin{equation}
\bar{H}_E(\mathbf{r},\mathbf{p}) =
\frac{1}{\kappa(\mathbf{r},\mathbf{p})} (H(\mathbf{r},\mathbf{p}) - E),
\label{r4}
\end{equation}
where the (conserved) value of $\bar{H}_E$ along the trajectory is
$\bar{E} = 0$.

To prove this fact, we need only demonstrate $\dot{X} = \{ X,
\bar{H}_E \}$, where $X(\mathbf{r},\mathbf{p})$ is an arbitrary phase
space quantity, $\{ \;, \; \}$ denotes the Poisson bracket, and the
dot denotes differentiation with respect to $\tau$.  To this end,
\begin{align}
 \{ X, \bar{H}_E \} & =  \{ X, \kappa^{-1} (H - E) \} \\
&= \{ X, \kappa^{-1} \}(H - E)  + \kappa^{-1} \{ X, H - E
\} \\
&= \kappa^{-1} \{ X, H \} = \kappa^{-1} X' = \dot{X},
\end{align}
where the second to last equality utilizes
\begin{equation}
X' = \{ X, H \},
\end{equation}
and the final equality uses Eq.~(\ref{r3}).

\subsection{Trajectory reparameterization for Lagrangian systems}

Suppose $\mathbf{R}(s)$ is a particular trajectory for the
$s$-independent Lagrangian $L(\mathbf{r},d\mathbf{r}/ds)$.  We again
let $E$ equal the value of the corresponding Hamiltonian along the
trajectory.  We also define a new parameterization $\tau$ by
Eq.~(\ref{r3}), except that $\kappa =\kappa (\mathbf{r},\mathbf{r}')$
is expressed as a function of $\mathbf{r}'$ rather than $\mathbf{p}$.
Then the trajectory $\tilde{\mathbf{R}}(\tau)$, referred to the new
parameterization, is a trajectory of the Lagrangian
\begin{equation}
\bar{L}_E(\mathbf{r},\dot{\mathbf{r}}) =
\frac{1}{\tilde{\kappa}(\mathbf{r},\dot{\mathbf{r}})} (\tilde{L}(\mathbf{r},\dot{\mathbf{r}}) + E),
\label{r5}
\end{equation}
where $\tilde{L}$ and $\tilde{\kappa}$ are $L$ and $\kappa$
re-expressed as functions of
\begin{equation}
\dot{\mathbf{r}} =\frac{1}{\kappa} \mathbf{r}'.
\label{r15}
\end{equation}
To prove this, we recall the Legendre transforms
\begin{align}
H &= \langle \mathbf{p}, \mathbf{r}' \rangle - L, \label{r17} \\
\bar{H}_E &= \langle \mathbf{p}, \dot{\mathbf{r}} \rangle -
\bar{L}_E, \label{r18}
\end{align}
where the momentum $\mathbf{p}$ is
\begin{equation}
\mathbf{p} = \partial L/ \partial \mathbf{r}' = \partial \bar{L}_E/ \partial \dot{\mathbf{r}}.
\label{r19}
\end{equation}
Substituting Eqs.~(\ref{r17}) and (\ref{r18}) into Eq.~(\ref{r4})
yields Eq.~(\ref{r5}).

\subsection{Reparameterizing stationary curves of the average
  Lagrangian shear}

\label{sec:ReparameterizingShear}

Defining the Lagrangian $L = p$, Eq.~(\ref{r7}) yields
\begin{equation}
L(\mathbf{r}, \mathbf{r}') = 
\frac{\langle \mathbf{r}',  D(\mathbf{r}, \theta) \; \mathbf{r}' \rangle}
{\sqrt{ \langle \mathbf{r}',\mathbf{r}' \rangle
\langle \mathbf{r}', C(\mathbf{r}, \theta) \;  \mathbf{r}' \rangle}}.
\label{r16}
\end{equation}
Re-expressing this in terms of the transformed derivative
$\dot{\mathbf{r}}$, Eq.~(\ref{r15}), yields
\begin{equation}
\tilde{L}(\mathbf{r},\dot{\mathbf{r}}) = 
\frac{\langle \dot{\mathbf{r}},  D \; \dot{\mathbf{r}} \rangle}
{\sqrt{ \langle \dot{\mathbf{r}},\dot{\mathbf{r}} \rangle
\langle \dot{\mathbf{r}}, C \;  \dot{\mathbf{r}} \rangle}},
\end{equation}
which has the same functional form as Eq.~(\ref{r16}),
since the $\mathbf{r}$ derivative occurs at the same order (two) in both the
numerator and denominator.  We now choose
\begin{equation}
  \tilde{\kappa}(\mathbf{r},\dot{\mathbf{r}}) = 
  \frac{1}{\sqrt{ \langle \dot{\mathbf{r}},\dot{\mathbf{r}} \rangle
      \langle \dot{\mathbf{r}}, C \;  \dot{\mathbf{r}} \rangle}},
\end{equation}
so that Eq.~(\ref{r5}) becomes
\begin{equation}
\bar{L}_E(\mathbf{r},\dot{\mathbf{r}}) =
\langle \dot{\mathbf{r}},  D \; \dot{\mathbf{r}} \rangle + E
\sqrt{ \langle \dot{\mathbf{r}},\dot{\mathbf{r}} \rangle
\langle \dot{\mathbf{r}}, C \;  \dot{\mathbf{r}} \rangle}.
\end{equation}

Equation~(\ref{r19}) implies $\langle \mathbf{p}, \mathbf{r}' \rangle
= \langle \mathbf{r}' , \partial L/ \partial \mathbf{r}' \rangle$,
which is just the derivative of $L$ in the direction of increasing
radius $|\mathbf{r}'|$ in the $\mathbf{r}'$-space.  Investigating
Eq.~(\ref{r16}), however, we see that the dependence of $L$ on
$\mathbf{r}'$ is solely through the angle $\theta$ of $\mathbf{r}'$
and not $|\mathbf{r}'|$.  Thus, $\langle \mathbf{p}, \mathbf{r}'
\rangle = 0$.  Furthermore, since $\mathbf{r}'= \kappa
\dot{\mathbf{r}}$, $\langle \mathbf{p}, \dot{\mathbf{r}} \rangle = 0$
as well.  Thus, Eqs.~(\ref{r17}) and (\ref{r18}) imply
\begin{equation}
H = L, \quad \quad \bar{H}_E = \bar{L}_E.  
\label{r20}
\end{equation}
Since $\bar{E} = 0$, the value of $\bar{L}_E$ along the stationary
trajectory $\mathbf{R}(\tau)$ also vanishes.  These results were
obtained by FBH~\cite{Farazmand14} through an alternative derivation.

\begin{theorem}
  Every stationary curve $\gamma$ of the average Lagrangian shear
  $P[\gamma]$, Eq.~(\ref{r6}), is also a stationary curve of
\begin{align}
  \bar{P}_E[\gamma] 
  &= \frac{1}{\tau_f} \int_0^{\tau_f}
  \bar{L}_E(\mathbf{r}, \dot{\mathbf{r}}) d \tau \\
  &= \frac{1}{\tau_f} \int_0^{\tau_f}
\left( \langle \dot{\mathbf{r}},  D \; \dot{\mathbf{r}} \rangle + E
\sqrt{ \langle \dot{\mathbf{r}},\dot{\mathbf{r}} \rangle
\langle \dot{\mathbf{r}}, C \;  \dot{\mathbf{r}} \rangle} \right) d \tau,
\label{r21}
\end{align}
where $E$ is the (conserved) value of the Lagrangian shear $p$ along
$\gamma$.  The value of $\bar{L}_E$ along $\gamma$ is also conserved
and equal to 0.  Note that the parameterization
$\bar{\mathbf{R}}(\tau)$ of $\gamma$ that makes $\bar{P}_E$ stationary
is typically different from the parameterization $\mathbf{R}(s)$ of
$\gamma$ that makes $P$ stationary.
\end{theorem}

\subsection{Perfect shearless fronts}

\label{sec:ShearlessFronts}

Equations~(\ref{r16}) and (\ref{r20}) show that $E = 0$ is equivalent to
$\langle \dot{\mathbf{r}}, D \dot{\mathbf{r}} \rangle = 0$.
Furthermore, one can show that any curve satisfying $\langle
\dot{\mathbf{r}}, D \dot{\mathbf{r}} \rangle = 0$ is a stationary
curve of Eq.~(\ref{r21}).  Thus, we have the following.
\begin{theorem}
  A curve $\gamma$ satisfying $\langle \dot{\mathbf{r}}, D \;
  \dot{\mathbf{r}} \rangle = 0$, or equivalently $p=0$, is a
  stationary curve of
\begin{equation}
  \bar{P}_0[\gamma] 
 = \frac{1}{T} \int_0^T
\langle \dot{\mathbf{r}},  D \; \dot{\mathbf{r}} \rangle d \tau.
\end{equation}
\label{thm:ShearlessFronts}
\end{theorem}
We call such curves \emph{perfect shearless fronts} (analogous to the
perfect shearless barriers of FBH~\cite{Farazmand14}), or just
shearless fronts for short.  They are the focus of the remainder of
this article.

Note that since $D$ is symmetric and traceless [see Eq.~(\ref{r22})],
it is like a pseudo-Riemannian metric tensor; ``pseudo'' because it
has signature $(-1, 1)$.  However, the resulting metric fails to be
given purely by a bilinear form, since $D$ itself depends on $\theta$.
Such a metric is called a (pseudo-)Finsler metric.  Perfect shearless
fronts are thus light-like, or null, geodesics of this pseudo-Finsler
metric.

\subsection{Computing perfect shearless fronts}

\label{sec:computingSF}

In Sects.~\ref{sec:variation}, \ref{sec:ReparameterizingShear}, and
\ref{sec:ShearlessFronts}, fronts were viewed as curves in the
$xy$-position space for the purpose of the variational problem.
Referring now to the full $xy\theta$-phase space, a shearless front
$(\mathbf{r}(\tau), \theta(\tau))$ satisfies both the front
compatibility criterion Eq.~(\ref{r26}) and $\langle \dot{\mathbf{r}},
D \dot{\mathbf{r}} \rangle = 0$.  The latter implies that $\langle
\dot{\mathbf{r}}, C \Omega \dot{\mathbf{r}} \rangle = 0$, and hence
that $\dot{\mathbf{r}}$ is an eigenvector of $C(\mathbf{r},\theta)$.
One possible approach to finding the shearless fronts, then, is to
integrate an eigenvector field of $C(\mathbf{r},\theta)$.
Let $\hat{\xi}(\mathbf{r},\theta)$ denote a choice of unit eigenvector
over $xy\theta$-space.  Taking $\dot{\mathbf{r}}$ proportional to
$\hat{\xi}$, the front compatibility criterion Eq.~(\ref{r26}) becomes
\begin{equation}
\chi(\mathbf{r},\theta) = 0,
\label{r9}
\end{equation}
where
\begin{equation}
\chi(\mathbf{r},\theta) = \mathbf{\hat{n}}(\theta) \cdot
\hat{\mathbf{\xi}}(\mathbf{r},\theta)
= (\sin \theta \mathbf{\hat{x}}
-\cos \theta \mathbf{\hat{y}})
\cdot \hat{\mathbf{\xi}}(\mathbf{r},\theta).
\end{equation}
This quantity is zero at $(\mathbf{r},\theta)$ if and only if
$\mathbf{\hat{g}}(\theta) = \pm
\hat{\mathbf{\xi}}(\mathbf{r},\theta)$.  If the choice of
$\hat{\xi}(\mathbf{r},\theta)$ were smooth everywhere, then
Eq.~(\ref{r9}) would define a smooth two-dimensional constraint
manifold in $xy\theta$-space.  In general, however, there is no
continuous choice of $\hat{\xi}(\mathbf{r},\theta)$ available over the
entire phase space.  The continuity of $\hat{\xi}(\mathbf{r},\theta)$
can break down at points in $xy\theta$-space where
$C(\mathbf{r},\theta)$ has degenerate eigenvalues, and hence where
there is not a unique eigendirection.  Since degeneracy of real
symmetric two-by-two matrices is a codimension-two criterion, the
degeneracy points generically occur along curves in $\mathbf{r}
\theta$-space.  Even if we remove these degenerate curves from
consideration, $\hat{\xi}(\mathbf{r},\theta)$ can not in general be
chosen continuously on the remainder of $xy\theta$-space.  This is due
to the fact that $\hat{\xi}(\mathbf{r},\theta)$ may obtain an overall
rotation by $\pi/2$ when transported along a loop encircling a
degenerate curve.

Rather than finding the eigenvectors explicitly, an improved approach
is to use the constraint equation
\begin{equation}
\psi(\mathbf{r},\theta) = 0,
\label{r13}
\end{equation}
where
\begin{align}
\psi(\mathbf{r},\theta) &= \mathbf{\hat{n}}(\theta) \cdot
C(\mathbf{r},\theta) \mathbf{\hat{g}}(\theta)
\\
&=   (\sin \theta \mathbf{\hat{x}}
-\cos \theta \mathbf{\hat{y}})   \cdot
C(\mathbf{r},\theta) 
(\cos \theta \mathbf{\hat{x}}
+\sin \theta \mathbf{\hat{y}}).
\end{align}
This quantity is zero at $(\mathbf{r},\theta)$ if and only if
$\mathbf{\hat{g}} (\theta)$ is \emph{any} eigenvector of
$C(\mathbf{r},\theta)$, not just a particular choice
$\hat{\xi}(\mathbf{r},\theta)$ as above.  The constraint surface
Eq.~(\ref{r13}) can thus be understood as the union of the two
surfaces defined by Eq.~(\ref{r9}) for the two choices of eigenvector
$\hat{\mathbf{\xi}}$.  The constraint surface Eq.~(\ref{r13}) is a
smooth surface without boundary (though potentially with
self-intersections), which we call the \emph{shearless surface}.

A perfect shearless front, lifted into $xy\theta$-space via the front
compatibility condition, must lie within the shearless surface.  A
perfect shearless front is thus an integral curve of a vector field
tangent to the shearless surface.  This vector field, normalized to
unity, is
\begin{equation}
\begin{aligned}
\frac{d \mathbf{r} }{d \lambda} &= \frac{a}{\sqrt{a^2+b^2}} \mathbf{\hat{g}}, \\
\frac{d \theta}{d \lambda} &= \frac{b}{\sqrt{a^2+b^2}}, 
\end{aligned}
\label{r1}
\end{equation}
where
\begin{equation}
\begin{aligned}
a &= - \frac{d}{d \theta} \psi, \\
b &= \mathbf{\hat{g}} \cdot {\bm \nabla} \psi.
\end{aligned}
\label{r12}
\end{equation}
An integral curve of (\ref{r1}) will lie within the shearless surface
so long as its initial condition does.  Note that these vector fields
do not require us to find the eigenvectors of $C(\mathbf{r},\theta)$.

We shall also need the vector field that is everywhere orthogonal to
Eq.~(\ref{r1}), but still tangent to the shearless surface.  This
field is obtained by replacing $\hat{\mathbf{g}}$ by
$\hat{\mathbf{n}}$ in Eqs.~(\ref{r1}) and (\ref{r12}), i.e., 
\begin{equation}
\begin{aligned}
\frac{d \mathbf{r} }{d \kappa} &= \frac{a}{\sqrt{a^2+c^2}} \mathbf{\hat{n}}, \\
\frac{d \theta}{d \kappa} &= \frac{c}{\sqrt{a^2+c^2}}, 
\end{aligned}
\label{r23}
\end{equation}
where $\kappa$ parameterizes the curve and
\begin{equation}
\begin{aligned}
a &= - \frac{d}{d \theta} \psi, \\
c &= \mathbf{\hat{n}} \cdot {\bm \nabla} \psi.
\end{aligned}
\end{equation}

\subsection{Burning Lagrangian coherent structures (bLCSs) as
  shearless fronts of maximal normal repulsion}

\label{sec:NormalRepulsion}

The normal repulsion $\rho$ of a front element $(\mathbf{r},\theta)$
is given by Eq.~(\ref{r29}).  On the shearless surface, where
$\hat{\mathbf{n}}$ is an eigenvector of $C$, this simplifies to
\begin{align}
  \rho(\mathbf{r},\theta) &= \sqrt{\hat{\mathbf{n}} \cdot
    C(\mathbf{r},\theta) \hat{\mathbf{n}}} = \sqrt{\lambda_\perp},
\end{align}
where $\lambda_\perp$ is the eigenvalue of $C$ normal to the tangent.
We now define the \emph{burning} finite-time Lyapunov exponent (bFTLE)
field as
\begin{equation}
\Lambda = \frac{1}{2T}\log(\lambda_\perp) =
\frac{1}{T}\log(\rho). 
\label{r24}
\end{equation}
Note that the specification of the eigenvalue $\lambda_\perp$ is based
on the orientation of its eigenvector (normal to the front) rather
than its magnitude; thus $\lambda_\perp$ need not be the
\emph{largest} eigenvalue.  

Since the bFTLE is defined on the two-dimensional shearless surface,
we have overcome part of the challenge of dimensions highlighted in
Sect.~\ref{sec:intro}.  In particular, we can plot and visually
interpret images of the bFTLE (Fig.~\ref{fig:FTLET3}a).  In some cases
a bLCS follows a ridge of the bFTLE field.  However, we see no reason
why a ridge of the bFTLE field need satisfy the front-compatibility
criterion, and so we opt not to compute bFTLE ridges.  Again, we see
the advantage of using the shearless fronts, for which the problem of
defining the bLCS becomes the problem of selecting the ``best'' curve
from the one-parameter family of shearless fronts.  We choose the
repelling bLCS to be that shearless front that (locally) maximizes
the average normal repulsion
\begin{equation}
R[\gamma] = \frac{1}{L}\int_\gamma \sqrt{\hat{\mathbf{n}} \cdot
  C(\mathbf{r},\theta) \hat{\mathbf{n}}} \; d\ell,
\label{r25}
\end{equation}
where $d \ell$ is the $xy$-euclidean line element and $L$ is the
euclidean length of the curve.  Similarly, the attracting bLCS is that
shearless front that (locally) minimizes the average normal repulsion.

There is still a question of what interval length $L$ to choose.  In
keeping with the intuitive idea of following an bFTLE ridge, we shall
adopt the approach of Refs.~\cite{Haller11,Farazmand12} and restrict
the length of a shearless front to those regions of the shearless
surface where the second derivative of the bFTLE field in the
direction normal to the front, given by Eqs.~(\ref{r23}), is negative.
In practice, the exact length used should make little difference, so
we use the curvature criterion to tell us approximately where to cut
the shearless curves.

\subsection{Cusps in the shearless fronts}

The shearless fronts, and hence a bLCS, may exhibit cusps when plotted
in $xy$-space.  This is one consequence of the 3D phase-space geometry
that does not occur for traditional LCSs in passive advection.  Cusps
have already been recognized as important features of BIMs for
time-independent and time-periodic flows (see
Fig.~\ref{fig:WindySingleVortex}a); cusps change the one-way barrier
direction of the BIMs.  Here, we determine when a cusp occurs along a
shearless front.  First, we consider the fold of the shearless surface
under the projection from the $xy\theta$-space to the $xy$-space; a
fold is where two branches of this projection meet.  Since a shearless
front is smooth on the shearless surface, if it remains on one branch
of this surface, then its projection to $xy$-space is also smooth.
Thus, a cusp of a shearless front can only occur at a fold of the
shearless surface.  What is less obvious, but nevertheless true, is
that every time a shearless front passes around a fold, its
$xy$-projection forms a cusp.  To prove this, consider a shearless
front $(\mathbf{r}(s),\theta(s))$ as it passes around a fold.
Locally, $\mathbf{r}$ can be expressed as a function of $\theta$,
which by Eqs.~(\ref{r1}) satisfies
\begin{equation}
\frac{d \mathbf{r} }{d \theta} = \frac{a}{b} \mathbf{\hat{g}}.
\label{r27}
\end{equation}
At the fold of the shearless surface, the 3D normal vector to the
surface, equal to $({\bm \nabla} \psi, d \psi/ d \theta)$, has no
component in the $\theta$ direction.  That is, $d \psi/ d \theta = 0$.
This implies that at a fold, $a = 0$ by Eq.~(\ref{r12}), and hence ${d
  \mathbf{r} }/{d \theta} = 0$, by Eq.~(\ref{r27}).  More precisely,
one sees that, at a fold, ${d \mathbf{r} }/{d \theta}$ transitions
from pointing in one direction to pointing in the opposite direction.
This is exactly the criterion for forming a cusp---the curve
$\mathbf{r}(\theta)$ moves forward and then backs up.  In summary,
then, a cusp occurs at exactly those points where a shearless front
passes transversely through a fold in the shearless surface.

\section{A test case}
\label{sec:DoubleVortex}

\subsection{Double-vortex in a windy channel}

\begin{figure}
\centering
\includegraphics[width=\linewidth]{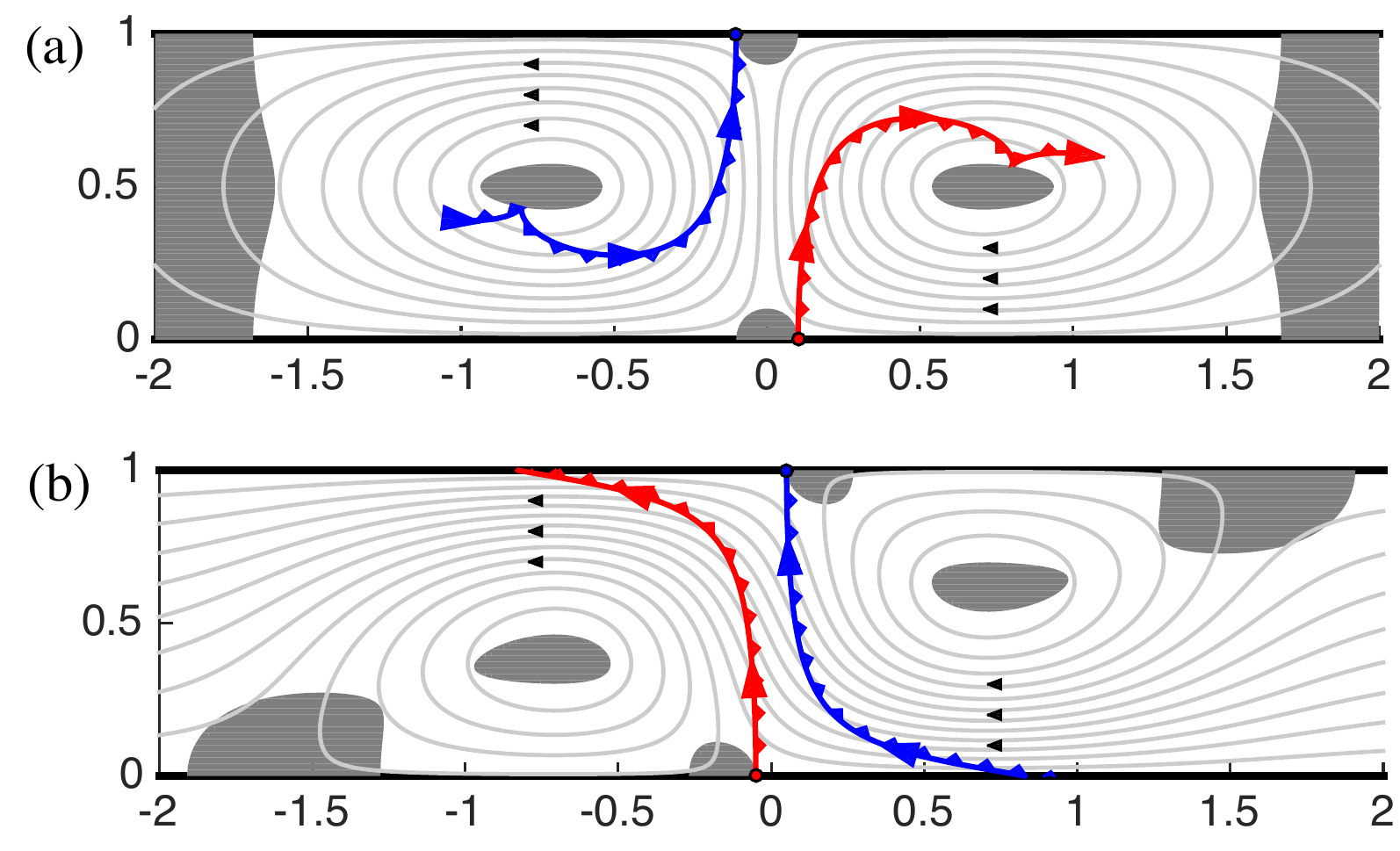}
\caption{\label{fig:WindySingleVortex} (color online) Double vortex
  channel flow with vortex strength $\gamma = -1.0$ and burning speed
  $v_0 = 0.1$.  a) no wind, $v_w = 0$; b) $v_w = -0.15$.  The
  streamlines of the flow are illustrated by grey curves.  The grey
  regions are ``slow zones'', where the fluid speed $u$ is less than
  $v_0$.  The blue and red curves are stable and unstable BIMs
  respectively.  The BIMs begin at burning fixed points on the channel
  boundary.  For plot a, the BIMs have cusps where they intersect the
  central slow zones.  For plot b, the BIMs span the channel without
  forming a cusp.  The arrows normal to the BIMs illustrate the
  burning direction.  
}
\end{figure}

As an example, consider an infinitely long channel flow containing two
side-by-side, counter-rotating vortices plus a spatially uniform
``wind'' flowing from right to left.  The streamfunction is
\begin{equation}
\psi(x,y) = (\gamma / \pi) x \exp(-x^2) \sin(\pi y) + v_w y,
\end{equation}
where $\gamma$ is the vortex strength and $v_w$ is the wind speed
(Fig.~\ref{fig:WindySingleVortex}).  Throughout our analysis we use
$\gamma = -1.0$, $v_w = -0.15$, and burning speed $v_0 = 0.1$.  The
channel width spans $y = 0$ to $y = 1$.  We choose this flow because
it models what can be realized in the laboratory of Tom Solomon
\cite{Paoletti05,Paoletti05b,Paoletti06,Schwartz08,Boehmer08,OMalley11}.
Furthermore, the BIM spans the channel without a cusp, which
simplifies our analysis.
Though in this paper we keep the wind speed constant, a future study
will consider time-varying wind.  Importantly, such time-varying wind
can be realized in the Solomon lab~\cite{Gowen14}.

For a steady wind, there are exactly two hyperbolic advective fixed
points of the flow, assuming the wind is not too large.  In
Figs.~\ref{fig:WindySingleVortex}a and \ref{fig:WindySingleVortex}b,
these fixed points lie within the gray ``semicircles'' attached to the
top and bottom channel wall, near the channel midpoint.  The gray
regions are ``slow zones'', regions where the fluid speed $u$ is less
than the burning speed $v_0$.  The bottom fixed point in these figures
generates an unstable manifold transverse to the channel wall, while
the top point generates a stable manifold.  The burning dynamics
induces the bottom advective fixed point to split into two SSU burning
fixed points, one at each intersection point between the boundary of
the slow zone and the channel wall.  (Only the rightmost burning fixed
point is shown in Fig.~\ref{fig:WindySingleVortex}.)  There are also
two SUU points near the bottom of the channel that do not concern us
here.  The right-burning SSU fixed point generates an unstable BIM,
shown in red in Fig.~\ref{fig:WindySingleVortex}.  Similarly, the
advective fixed point at the top of the channel splits into two SUU burning
fixed points on the upper channel boundary.  The leftmost of these is
shown in Fig.~\ref{fig:WindySingleVortex}, along with the attached
stable BIM in blue.

For weak or no wind, each BIM forms a cusp before crossing the channel
(Fig.~\ref{fig:WindySingleVortex}a.)  As the wind speed increases
above $v_w = v_0$, each BIM crosses the channel without forming a cusp
(Fig.~\ref{fig:WindySingleVortex}b.)  The process by which the BIM
attaches to the boundary is described in detail in
Ref.~\cite{Mahoney15b}.  For $v_w > v_0$, each BIM divides the channel
in two.  If the entire region left of the unstable BIM in
Fig.~\ref{fig:WindySingleVortex}b is burned, then it will remain
burned for all time, forming a frozen front along the
BIM~\cite{Mahoney15b,Gowen14}.  In contrast, for $v_w < v_0$, any
initially burned region will eventually propagate arbitrarily far down
the channel to both the left \emph{and} the right.  Thus, the
existence of an unstable BIM that traverses the channel with no cusp
has a very clear experimental signature.

A stable BIM traversing the channel without a cusp
(Fig.~\ref{fig:WindySingleVortex}b) forms a kind of basin boundary.
An initial point stimulation left of the stable BIM will eventually
grow to become a frozen front.  However, a stimulation right of the
stable BIM will eventually fill up the entire channel.  In the
remainder, we shall focus on the stable BIM.

\subsection{The bLCS for a steady wind---short integration time}

\label{sec:short}

\begin{figure}
\centering
\includegraphics[width=1.0\linewidth]{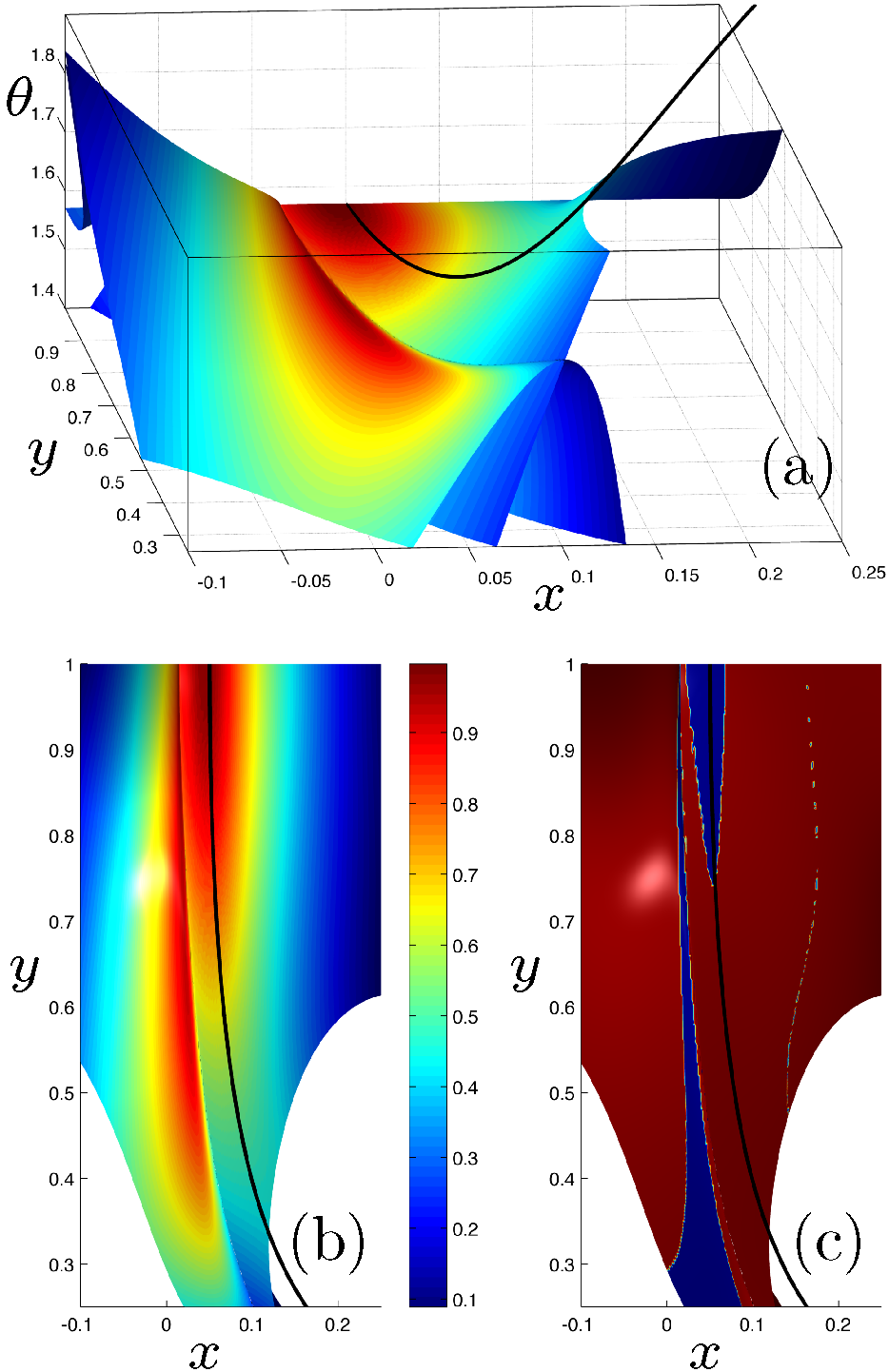}

\caption{\label{fig:FTLET3} (color online) a) The two-dimensional
  shearless ($\psi = 0$) surface embedded in the three-dimensional
  phase space.  The coloring represents the two-dimensional bFTLE
  field on this surface.  The Cauchy-Green integration time is $T =
  3$.  The black curve is the stable BIM attached to the rightward
  propagating burning fixed point at the top of the channel.  The
  computation of the isosurface is restricted to the computation box
  shown.  b) A view of the bFTLE field on the shearless surface
  projected onto the $xy$-plane.  The two white patches on the bottom
  left and right are where the shearless surface has exited the
  computation box. c) The sign of the bFTLE curvature normal to the
  shearless fronts (see Eq.~(\ref{r23}).  Blue is negative and red is
  positive.  }
\end{figure}

We now apply the method of Sect.~\ref{sec:CoreAnalysis} to the steady
wind case ($v_w = -0.15$) with the relatively short Cauchy-Green
integration time $T=3$.  We should find a bLCS that follows the
initial length of the time-independent BIM.  Before finding the bLCS,
however, it is helpful to visualize the shearless surface.
Fig.~\ref{fig:FTLET3}a shows the surface in 3D.  For comparison, the
stable BIM is shown in black.  Near the burning fixed point, the BIM
lies near (though not precisely within) the shearless surface.  This
is an important first check on the validity of our technique; if the
BIM were far from the shearless surface, we would have no chance of
recovering it.

At the bottom center of Fig.~\ref{fig:FTLET3}a is a ``pitchfork''
intersection between two branches of the surface.  The
differentiability of the $\psi$ field implies that the intersection
curve between these two branches must be a shearless front, i.e. a
trajectory of Eqs.~(\ref{r1}).  Thus, no shearless fronts may pass
through this intersection.

The coloring in Figs.~\ref{fig:FTLET3}a and \ref{fig:FTLET3}b
represents the bFTLE field Eq.~(\ref{r24}).  Notice that the BIM
follows a bFTLE ridge at the top of Fig.~\ref{fig:FTLET3}b, another
critical check on our technique.  Curiously, there is a second bFTLE
ridge on the left in Fig.~\ref{fig:FTLET3}b.  This ridge lies on a
separate branch of the pitchfork structure, as seen in
Fig.~\ref{fig:FTLET3}a.

We next identify those regions where the second derivative of the
bFTLE field normal to the shearless curves is negative.
Figure~\ref{fig:FTLET3}c illustrates the sign of this curvature, with
blue representing negative and red representing positive.  There are
two regions of negative curvature, one for each of the bFTLE ridges.
We focus on the rightmost region associated with the BIM.

\begin{figure}
\centering
\includegraphics[width=1.0\linewidth]{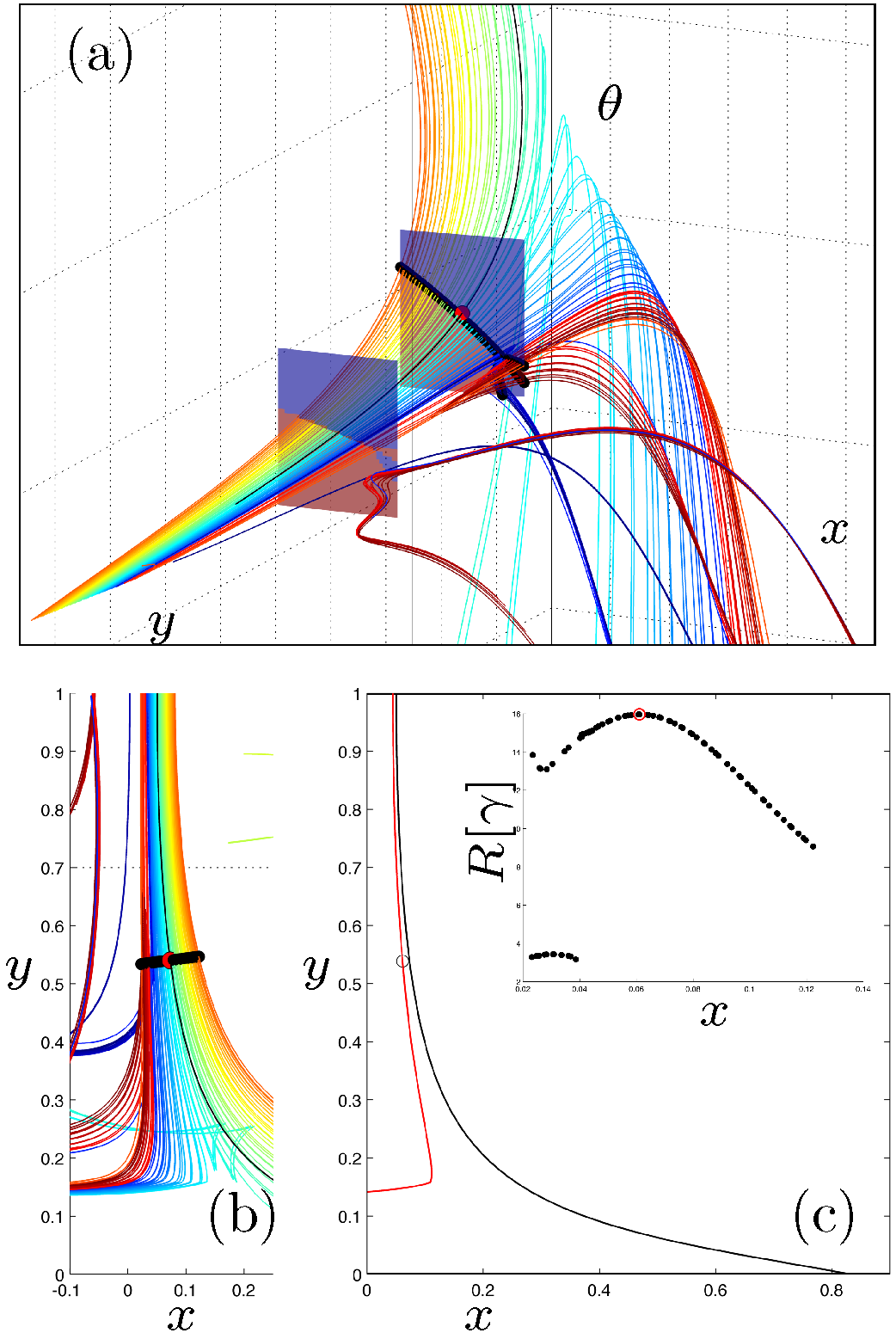}

\caption{\label{fig:ShearlessCurves3D} (color online) 
  a) Shearless fronts computed for Cauchy-Green integration time $T =
  3$.  The initial points (bold) of the trajectories were computed by
  intersecting a plane transverse to the BIM (at the red dot) with the
  shearless surface.  The BIM is the thin black curve.  b) The
  shearless fronts are shown projected onto the $xy$-plane.  c-inset)
  Plot of the average normal repulsion, Eq.~(\ref{r25}), as a function
  of the $x$-coordinate of the initial point for each trajectory in
  parts a and b.  A smooth maximum is marked in red and determines
  which shearless front is the bLCS. c) The bLCS is plotted in red
  next to the BIM in black.  The open circle marks the initial
  condition for generating the bLCS.}
\end{figure}

We next compute shearless fronts, targeting the bFTLE ridge on the
right.  We select initial conditions using the intersection of the
shearless surface with a plane normal to the BIM, yielding the bold
line of initial points in Fig.~\ref{fig:ShearlessCurves3D}a~\cite{Note1}. 
%
%
We integrate Eqs.~(\ref{r1}) upward toward the top of the channel and
downward toward the bottom.  Figures~\ref{fig:ShearlessCurves3D}a and
\ref{fig:ShearlessCurves3D}b show these shearless fronts in 3D and 2D,
respectively.

We then compute the average normal repulsion, Eq.~(\ref{r25}), along
the segments of the shearless curves above the dashed line $y = 0.7$
in Fig.~\ref{fig:ShearlessCurves3D}b.  This line is chosen to lie
close to the bottom of the negative curvature region in
Fig.~\ref{fig:FTLET3}c.  The resulting average normal repulsion is
shown in the inset of Fig.~\ref{fig:ShearlessCurves3D}c.  It exhibits
a clear local maximum, which selects the red shearless curve isolated
in Fig.~\ref{fig:ShearlessCurves3D}c.  This curve is the bLCS, and it
faithfully follows the initial length of the BIM, as anticipated,
before eventually deviating markedly from it.

\subsection{The bLCS for a steady wind---long integration
  time}

\label{sec:long}

\begin{figure}
\centering
\includegraphics[width=1.0\linewidth]{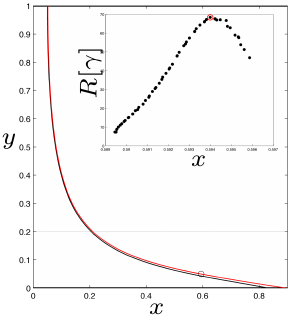}
\caption{\label{fig:T6} (color online) The bLCS (red) and the BIM
  (black) for Cauchy-Green integration time $T = 6$.  The open circle
  on the bLCS marks the initial point used to integrate the shearless
  front.  (The neighboring shearless fronts are not shown.)  The inset
  shows the average repulsion on a shearless front as a function of
  the $x$-coordinate of its initial point.  The average was computed
  along the shearless fronts above the dashed line at $y=0.2$.}
\end{figure}

We extract a bLCS for the same dynamics as in Sect.~\ref{sec:short}
but for a longer Cauchy-Green integration time of $T = 6$.  The result
is summarized in Fig.~\ref{fig:T6}.  Clearly the bLCS (red) closely
tracks the BIM (black), considerably longer than for the $T = 3$ case.
In fact, the agreement is quite good all the way to the bottom of the
channel.  This is exactly what we expect to see as the integration
time $T$ is increased.

\section{Conclusion}

We have developed a technique to extract maximally repelling (or
attracting) fronts, called burning Lagrangian coherent structures
(bLCSs) for the finite-time dynamics of fronts evolving within a
flowing fluid.  We verified that this technique returns the burning
invariant manifolds (BIMs) for the case of a time-independent flow.
Clearly more work is needed to establish the relevance of these bLCSs
to time-periodic and, most importantly, to time-aperiodic flows.  This
will be pursued in a separate publication.

\acknowledgments

This work was supported by the US National Science Foundation under
grant CMMI-1201236.  The authors gratefully acknowledge discussion
with Tom Solomon, George Haller, and Mohammad Farazmand.


%

\end{document}